\documentstyle[aps,epsf,a4]{revtex}
\title{ String formation and chiral symmetry breaking
in the heavy--light quark--antiquark system in QCD}
\author{Yu. A. Simonov}
\address{Institute of Theoretical and Experimental Physics, Moscow, Russia}
\author{J. A. Tjon}
\address{Institute for Theoretical
Physics, University of Utrecht, 3584 CC Utrecht, The Netherlands and
KVI, University of Groningen, 9747 AA Groningen, The Netherlands}
\date{\today}
\newcommand{\be}{\begin{equation}}
\newcommand{\ee}{\end{equation}}
\newcommand{\bea}{\begin{eqnarray}}
\newcommand{\eea}{\end{eqnarray}}

\def\ga{\mathrel{\mathpalette\fun >}}
\def\fun#1#2{\lower3.6pt\vbox{\baselineskip0pt\lineskip.9pt
\ialign{$\mathsurround=0pt#1\hfil
##\hfil$\crcr#2\crcr\sim\crcr}}}

\newcommand{\vex}{\mbox{\boldmath${\rm x}$}}
\newcommand{\vey}{\mbox{\boldmath${\rm y}$}}
\newcommand{\ver}{\mbox{\boldmath${\rm r}$}}
\newcommand{\vez}{\mbox{\boldmath${\rm z}$}}
\newcommand{\veal}{\mbox{\boldmath${\rm \alpha}$}}
\def\figlab#1{\label{fig:#1}}

\begin{document}
\maketitle

\begin{abstract}

The effective quark Lagrangian is written for a light quark in the
field of a static  antiquark, explicitly containing field correlators
as coefficient functions of products  of quark operators. At large $N_c$
the closed system of  equations for the  gauge--invariant quark
Green's function in the field of static source is examined
analytically. The
formation of the string connecting the light quark to the static
source is observed numerically. The scalar Lorentz nature of the
resulting confinement is shown to hold for the considered
case, implying chiral symmetry breaking. The resulting spectrum with
and without perturbative gluon exchanges is obtained numerically and
compared to the $B$ and $D$ meson masses and HQET.
\end{abstract}
\vspace {2cm}

 Pacs: {12.38Gc, 12.38Lg,12.39.Hg}

\newpage
\section{Introduction}

The nature of QCD string formation between static  sources was studied on the
lattice \cite{1,2} and analytically \cite{2}. From these investigations it was
shown that the string consists of a predominantly color--electric longitudinal field.
At the critical temperature $T_c$ this electric field disappears and
and at the same time
the deconfined phase with color--magnetic condensate sets in. This
effect was predicted theoretically in \cite{3} and also seen in lattice
measurements\cite{4}. At the same temperature Chiral Symmetry
Breaking  (CSB) for light quarks is found to disappear\cite{CG}, 
which indicates that there is
an    intimate connection between the string formation and CSB.
In the case of heavy quark systems CSB occurs due to the quark
mass and confinement can be described as the area law of the Wilson
loop. How CSB and confinement are explicitly realized for the light quark
system  and what equation describes its  dynamics is an interesting and 
open problem. 

It is the purpose of the present paper to study this issue in the
simplest dynamical example -- in the system of one light quark and a
heavy antiquark. This allows us to describe the dynamics of
light quark (its propagator) in a gauge--invariant manner, while
physically the light quark is expected to be  confined at the end of
a string connected to the static source.
Applying the formalism of field correlators (FC) \cite{DS,S1}, we
derive the effective quark Lagrangian, containing any number of
quark operators multiplied by  field correlators.
To proceed  further
one can use the limit of large $N_c$ and write down the Dyson-Schwinger
equations with the mass operator expressed through the Green's
function. The resulting equations are nonlocal and nonlinear.
It is not clear from the beginning how confinement and CSB would
manifest themselves in the solution of these equations. Some hint
was provided in Ref. \cite{S2} using a relativistic WKB analysis
\cite{PM}, where
it was shown that at large distances from the heavy source the
dynamics of the light quark is described by the  Dirac equation with a
scalar linear confining interaction, which leads to CSB.

In this paper we examine the properties of the Green's function of
the color singlet $q \bar Q$ system, where the antiquark is treated in
the static limit.
In section II we formulate the full form of nonlocal and nonlinear 
equations for the light quark propagator and its eigenfunctions  
and study the behaviour at all distances.
We also take into account both perturbative and
nonperturbative contributions to the interaction  kernel. As a
result our equations contain both a confining interaction and
color Coulomb part. 
Similar to the heavy quark situation we argue that the string formation 
for low angular momentum is of a color-electric nature. 
Moreover, the confinement of the 
light quark to the heavy one is shown to be of Lorentz scalar type.
In section III the resulting nonlinear equations are
studied numerically. The energy spectrum and the structure of the low lying
eigenfunctions are presented. We in particular study the $B$ and $D$ meson 
spectrum. We compare  them with experimental data for
$B$, $D$ mesons and results of other calculations, exploiting for this
purpose the expansion of Heavy Quark  Effective Theory (HQET).
 In section IV  the chiral condensate of the light quarks 
$\langle \bar q q\rangle $ is determined by taking the
limit of the Green's function $S(x,y)$, with
both $x$ and $y$ tending to zero. In this limit the heavy quark is
turned off and the condensate $\langle \bar q q\rangle \sim S(0,0)$ should 
have a value not depending on the presence of heavy quark.

\section{Derivation of equations}

\subsection{Dyson-Schwinger equations}

In this section we give an outline of the procedure to obtain the 
Dyson-Schwinger equations for the color white $q \bar Q$ system.
Our starting point is the gauge--invariant light quark Green's function $S(x,y)$
in the presence of a static heavy antiquark placed in the origin. In the static limit,
the heavy quark can be treated as an external source. Assuming the Euclidean metric 
and letting $T=(x-y)_4$,  the heavy antiquark propagator 
in the modified Fock-Schwinger gauge \cite{B} is proportional to the
parallel transporter,namely
\be
S_{\bar Q}(A) =  
 h({\bf x},{\bf y})\ P exp( i g \int_0^T A_4({\bf r}=0,\tau) d\tau)
\ee
with
$$
h({\bf x},{\bf y})=
\frac{i}{2} \delta^{(3)}({\bf x}- {\bf y}) [
 (1+\gamma_4)\ e^{-m_QT} \Theta(T) + ( T \rightarrow -T, \gamma_4 
 \rightarrow -\gamma_4)].
$$
This acts as a static source situated at the origin.
As a result the proper limit for $m_Q \rightarrow \infty$ of the 
Green's function of the  $q\bar Q$ system   can be defined as
\be
S(x,y) =  \langle \psi(x)\ \Pi (x,y)\ \psi^+(y) \rangle. 
 \label{3}
\ee
Here we have to average over the gluon fields $A$ and light quark fields 
$\psi$, while $\Pi$ contains the parallel transporters between the end points 
$$
\Pi(x,y) \equiv \phi({\bf x}, x_4; 0,x_4)\ \phi(0,x_4; 0, y_4)
\ \phi(0,y_4; {\bf y}, y_4)
$$
with
$$ \phi({\bf x},x_4;{\bf y},y_4) = P~exp ~i g \int^y_x  A_\mu dz_\mu$$.

The averaging over the gluon fields $A$  has to be done over the 
perturbative and nonperturbative gluon fields $a_\mu$ and $B_\mu$ 
contributions respectively,
where the total gluonic field $A_\mu=B_\mu+a_\mu$.
We now apply the method of field correlators (FC), which was developed
in a series of papers \cite{DS,S1} to derive the effective quark 
Lagrangian from QCD. 
Let us first consider the effects of averaging over nonperturbative (NP)
field $B_\mu$.  We may write for the  partition function
\be 
\langle P e^{g\int
\psi^+ \hat B(x)\psi dx}\rangle _B\equiv e^{L_{eff}} = P exp[
\sum_n\frac{ g^n}{n!}\int d^4x_1...  d^4x_nj(1)...  j(n) \langle\langle
B(1)...  B(n)\rangle\rangle ]
\label{1}
\ee
with $j(n)\equiv j_{\mu_n}(x_n)=\psi^+(x_n) \gamma_{\mu_n}\psi (x_n)$ and
$B(n)=B_{\mu_n}(x_n)$. 
To write the correlator $\langle\langle  B... B\rangle\rangle $ 
for the gauge--invariant situation corresponding to the color
white $q \bar Q$ system one can use the modified
Fock-Schwinger gauge \cite{B} to express  the correlator of
$B(x)$ through FC:
\be
N(1,...n)\equiv
 \langle\langle  B(1)... B(n)\rangle\rangle
\sim \int dx(1)... dx(n)
 \langle\langle  F(1)... F(n)\rangle\rangle .
 \label{2}
 \ee
The effective interaction kernel in  Eq. (\ref{1})  can now be used to write a
Dyson--Schwinger--type equation for the quark Green's function $S$.
To simplify matter, one can consider the large $N_c$ limit, in which
case the connected self--energy kernel $M(x,y)$ is obtained from
Eq. (\ref{1}) by replacing any pair of adjacent $\psi$--operators by
\be
\psi_{a\alpha}(x)\psi^+_{a\beta}(y)\to N_cS_{\alpha\beta} (x,y),
\label{4}
\ee
where $a\alpha$ are color and Lorentz index respectively (for details
of derivation see \cite{S2}). As a result one obtains the equation
for $S(x,y)$
\be
(-i \partial \!\!\!/_x ~ -im)S(x,y) -i\int M(x,z)S(z,y) d^4z= \delta^{(4)}(x-y),
\label{5}
\ee
where the kernel $M$ is expressed through $N$ as 
\bea
iM(x,y) &&= N^{(2)}_{\mu\nu}(x,y)\gamma_\mu S(x,y) \gamma_\nu +
\\
\nonumber
&&\sum^\infty_{n=3}\int d^4x_2... d^4x_{n-1} \gamma_{\mu_1}
S(x,x_2) \gamma_{\mu_2}...
\gamma_{\mu_{n-1}}S(x_{n-1},y)\gamma_{\mu_n}
N^{(n)}_{\mu_1...\mu_n}(x,x_2...x_{n-1}, y).
\label{6}
\eea
The system of equations (6-7) is exact in the large
$N_c$ limit and is well defined provided all NP correlators
$\langle\langle F(1)... F(n)\rangle\rangle$ are known.

Evidence has been found in recent accurate lattice calculations \cite{11}
of static potentials in different $SU(3)$ representations,  that 
the contributions of the higher correlators 
$\langle\langle F(1)... F(n)\rangle\rangle$ for $n>2$ to 
the planar Wilson loop are small.
In particular, these terms are found to contribute only around a few 
percent of the dominant Gaussian correlator \cite{12}.
Hence, the Gaussian Stochastic Model, based on the lowest correlator 
$\langle\langle F(1)F(2)\rangle\rangle$ is expected to be a good 
approximation. In view of this we assume in this paper that the Gaussian 
approximation holds and we keep similar as in Ref. \cite{S2} only the first term 
in Eq. (\ref{6}).  
One can parametrize the Gaussian correlator according to \cite{DS} as
  \be
 \langle\langle F(1) F(2)\rangle\rangle=
 \frac{1}{N_c}tr\langle
 F_{\mu\lambda}(x)\phi(x,0) F_{\nu\sigma} (0) \phi{(0,x)}\rangle =
D(x)(\delta_{\mu\nu} \delta_{\lambda\sigma}-\delta_{\mu\sigma}
\delta_{\nu\lambda})+\Delta^{(1)}_{\mu\lambda\nu\sigma},
\label{7}
\ee
where only $D(x)$ is responsible for confinement and it contributes to
string tension $\sigma$, while $\Delta^{(1)}$ is a full derivative. 
As a consequence, the latter contributes to the perimeter of Wilson loop and 
$\phi(x,0) =P~exp~ig \int^x_0  B_\mu dz_\mu$. From this we find, 
that the $N^{(2)}_{\mu\nu}$ can explicitly be
 written in the gauge \cite{B} as
 \be
N_{\mu\nu}^{(2)} (x,y) = (\delta_{\mu\nu} \delta_{ik}-\delta_{\mu
 k}\delta_{\nu i}) \int^x_0 du_i \alpha_\mu (u) \int^y_0 dv_k
 \alpha_\nu(v) D(u-v),
 \label{8}
 \ee
 where $\alpha_4(u)=1, \alpha_i(u)=\frac{u_i}{x_i}$. 

In Eq. (\ref{8}) only the nonperturbative confining piece of the
Gaussian correlator (\ref{7}) is retained, since the perturbative
part and $\Delta^{(1)}$ do not produce neither the string (confinement)
nor CSB \cite{S2}.  
In the case of a nonrotating string the terms in Eq.~(7)
with space components $\gamma_k$ are suppressed by powers of 
velocity of the endpoints of the string. In what follows we shall 
keep for simplicity only the component $N_{44}^{(2)}\equiv J(x,y)$ 
of $N_{\mu\nu}^{(2)}$.  Hence the kernel $M$ is
proportional to the FC of the color--electric field $E_i\equiv F_{i4}$.
It is the dominant part of the string. Color--magnetic components
are neglected in this first step and can be considered as a
correction.

In this way one arrives at the system of equations \cite{S2}
where we keep the same notation $M(x,y)$ for the retained piece of 
the kernel
\be
iM(x,y)=J(x,y) \gamma_4 S(x,y) \gamma_4
\label{9}
\ee
\be
(-i \partial \!\!\!/_x ~ -im) S(x,y) -i\int M(x,z) S(z,y) d^4 z 
=\delta^{(4)}(x-y).
\label{10}
\ee

\subsection{Partial wave reduced equations}

The kernel $ J(x,y)$ in  Eq. (\ref{9}) depends on the
time as $\frac{x_4-y_4}{T_g}$. Therefore in the limit of $T_g\to
0$  it becomes local in time.  As a
result, the Fourier--transform of $M$ in the fourth coordinate  does
not depend on the momentum $p_4$ for $T_g\to 0$. The corresponding
eigenfunctions $\psi_n(\vex, p_4)=\psi_n(\vex)$ and eigenvalues
$\varepsilon_n=\varepsilon_n(p_4)$ of the light quark in
the white light-heavy configuration can readily be obtained from
Eqs. (\ref{9}-\ref{10}) in the discussed instantaneous limit. We find
 \be
 (\frac{\veal}{i}\frac{\partial}
 {\partial\vex} + \beta m) \psi_n (\vec
 x) + \beta \int M(\vex, \vez) \psi_n(\vez) d^3\vez
=\varepsilon_n\psi_n(\vex)
\label{11}
 \ee
 \be
 M(\vex, \vez) = J(\vex, \vez)\beta \Lambda(\vex, \vez),\ \ \
 \Lambda(\vex, \vez) = \sum_k \psi_k(x)sign \varepsilon_k \psi^+_k(z)
 \label{12}
 \ee
 The spherical--spinor decomposition of $\psi_n$ for the total and orbital
angular momentum channel $j$, $l=j \pm \frac{1}{2}$,
 \be
 \psi_n(r)= \frac{1}{r}\left(
 \begin{array}{l}
 G_n(r)\Omega_{jlM}\\
 iF_n(r)\Omega_{jl'M}
 \end{array}\right), l'=2j-l
 \label{13}
 \ee
 yields equations for the partial waves
 \be
 \frac{dF_n}{dr}-\frac{\kappa}{r} F_n+(\varepsilon_n-m) G_n-
 M_{11} G_n-i M_{12} F_n=0
 \label{14}
 \ee
 \be
 \frac{dG_n}{dr}+\frac{\kappa}{r} G_n-(\varepsilon_n+m) F_n-
 M_{22} F_n+i M_{21} G_n=0
 \label{15}
 \ee
with $\kappa=\pm (j+\frac{1}{2})$. Clearly the kernels $ M_{ik}$ are
nonlocal in space, i.e.
 \be
  M_{ik} G_n=\int^\infty_0 M^{jj}_{ik}
 (r,r')~G_n(r') rr' dr'
\label{16}
\ee
with $M^{jj}_{ik}=\langle \Omega_{jl_iM}|M_{ik}| \Omega_{jl_kM} \rangle$
and $M_{ik}$ is given by Eq. (\ref{12}).

Equations (\ref{14}-\ref{15}) are invariant under the
transformation
 \be
  \varepsilon_n\to
-\varepsilon_n,\ \ \kappa\leftrightarrow-\kappa,\ \  G_n\leftrightarrow F_n
\label{17}
\ee
which also yields $M_{11}\leftrightarrow M_{22},
M_{12}\leftrightarrow-M_{21}.
$
The symmetry (\ref{17}) implies that the spectrum is symmetric in
$\varepsilon_n\leftrightarrow-\varepsilon_n$, which is a property of
a scalar interaction. The Lorentz scalar nature of the
confining interaction  has the nice feature that it doesnot lead to
instability problems in the Dirac equation \cite{T}.
Moreover, nontrivial solutions of Eqs. (\ref{14}-\ref{15}), if
they exist, signify spontaneous CSB.

We solve Eqs. (\ref{14}-\ref{15}), using the
relativistic WKB approximation for the kernel $ M_{ik}$. To
simplify the calculations the Gaussian form for
$D(x)$ was used
(since all observables are integrals of FC, its explicit form is not
essential  at large distances, provided the FC have a finite range $T_g$
and it yields the same value of
the string tension  $\sigma$)
\be
D(u)=D(0) exp
(-u^2/4T^2_g),\ \ D(0)=\frac{\sigma}{2\pi T^2_g}
\label{18}
\ee
with $\sigma=0.2~GeV^2,$ and $ T_g=0.25~fm$, taken  in accordance with
the lattice measurements \cite{4,GP}.

As the reference basis we take  the WKB solutions of the Dirac
equation for the local linear confining potential $\sigma r$,
 and the WKB computed kernels $\tilde M$ and $\tilde \Lambda(\vex, \vey)$
in Eq.~(\ref{12}) are determined by explicit summation over eigenstates.
It was checked by an independent calculation \cite{MNS} that the
relativistic WKB procedure yields eigenvalues of the linear potential
with accuracy better than one percent. 

The general structure of $M$ and $\tilde M$ can be derived from
Eqs. (\ref{12}-\ref{13}).
One can write $ M$ as a $4\times 4$ matrix as
follows \cite{S3}
\be
 M= M^{(0)} I+M^{(i)} \hat \sigma_i+ M^{(4)} \gamma_4
+M_\gamma^{(i)} \gamma_i,
\label{20}
\ee
where $\gamma_i, \gamma_4$ are usual Dirac matrices, $i=1,2,3$
and $\hat \sigma_i =
 \left (
 \begin{array}{ll}
 \sigma_i&0\\
 0&\sigma_i
 \end{array}\right )$.

 The same representation holds for the WKB approximated $\tilde M$.
 From the WKB analysis \cite{S2,S3} we find that
$\tilde M^{(0)}$ is the only growing kernel.
It behaves asymptotically as
 \be
 M^{(0)}(\vex,\vey) =\sigma \frac{|\vex +\vey|}{2} \tilde
 \delta^{(3)}(\vex-\vey)
 \label{21}
 \ee
 where $\tilde \delta^{(3)}(\ver)$ is a smeared $\delta$ --
 function with the range of nonlocality decreasing asymptotically
 with growing $|\vex|$, $|\vey|$. The term $M^{(i)}$ is
 proportional to the angular momentum $L$ and asymptotically it
 behaves as $O(1/x)$. One can also prove that $M^{(4)},
 M_\gamma^{(i)}$ do not grow at large $x,y$ \cite{S3}. Hence in all
 problems where large distances are dominant one can consider only
 the first term in Eq. (\ref{20}).
Using this approximation we get for the kernel (\ref{12}) 
\be
\tilde M = \tilde M^{(0)}( \vex, \vey) I = J( \vex, \vey) 
\tilde \Lambda( \vex, \vey) I 
\label{kern}
\ee
with
\be
J ( {\bf x}, {\bf y})
 = \sigma\frac{ {\bf x} {\bf y}}{\sqrt{\pi}}
 f ({\bf x}, {\bf y }),\ \ f
 ({\bf x}, {\bf y })
  =\int^1_0 ds \int^1_0 dt e^{-\frac{({\bf x}s-{\bf y}t)^2}{4T^2_g}}
\ee
and where $\tilde \Lambda$ is now a scalar quantity.

A curious feature of the considered equations is the way how the string
connecting the light quark to the source is being created. Actually, if
one takes only lowest partial waves inside the kernel $ M$ (i.e.
in $\Lambda(\vex, \vey)$, Eq. (\ref{12})), then the effective
potential in Eqs. (\ref{14}-\ref{15}) is not confining. If one
however sums up over all {\it{angular states and radial
excitations}} in $\Lambda$, then the resulting $\Lambda(x,y)$ is a
smeared $\delta $-- function leading to the quasilocal confining
kernel $\tilde M$.  E.g.  using eigenfunctions for the local case,
$\tilde \Lambda$ can be computed quasiclassically to be
\be
 \tilde \Lambda (\vex, \vey)= \frac{\sigma^2xy}{2\pi^2} \frac{K_1(\sigma
 \sqrt{xy}\sqrt{(x-y)^2+\theta^2xy)}}{\sqrt{(x-y)^2+\theta^2xy}},\ \ 
 \vex\vey=xy \cos \theta
 \label{22}
 \ee
 In Eq. (\ref{22}) one can clearly see that $\tilde
 \Lambda(x,y)$ is a normalized smeared $\delta$--function, with
 smearing radius in $|x-y|$ being $\frac{1}{\sigma\sqrt{xy}}$.  For
large distances it is nonvanishing only in the forward direction.

Insertion of this $\tilde \Lambda$ into $\tilde M$, Eq. (\ref{kern}),
 produces linear confinement  due to the kernel $\tilde \Lambda(\vex,
 \vey)$, as is given by Eq. (\ref{22}) (while simply averaging the
 contribution from each individual orbital in Eq. (\ref{12}) over
 the angle between $\vex$ and $\vey $ would produce no confinement at
 all).
The computed kernel
$\tilde M(x,y)$ turns out to be nonlocal, but very close to the
linear potential at large distances.  Indeed, the effective localized 
potential defined as
\be
V_{eff}(r)={\int M^{(0)}(r,x) rx{dx}}
\label{19}
\ee
approaches at very large distances, i.e. for $\sigma^{1/2} r>200$, a linear
dependence with a slope given by $\sigma$. However at shorter distances
$V_{eff}$ looks also linear over a relatively large region with a slope
almost the same as $\sigma$, reflecting the
presence of a small local curvature. In particular, we find that in the region
$5<\sigma^{1/2} r<20$ the effective potential can reasonable well be described
by $V_0(r)= 0.9 \sigma r -1.8\sigma^{1/2}$. In Fig.~1 are shown
the results up to $r=20$ in units of $\sigma^{1/2}$.

Since it is only the higher states and large distances which are
important in the creation of  this $\delta $ -- function--type behaviour
of $\Lambda (x,y)$,
and since the WKB method does well for high states and at large distances,
one can clearly conclude, that linear confinement should be obtained
if one sums over all exact solutions of Eqs.
(\ref{14}-\ref{15}).  Hence this should be a property of the
exact solution.

The property, that the kernel has a focussing effect in the forward direction
can be used to get a somewhat simpler form. For this purpose we may also use
\cite{S2}

\be
\tilde \Lambda (\vex, \vey)=
\frac{\sigma}{\pi^2 \sqrt{xy}}K_0(a) \delta(cos \theta_x -cos\theta_y) ,
\ \  a=\sigma \sqrt{xy} |x-y|
\label{23}
\ee

The eigenvalues and eigenfunctions in Eqs. (\ref{14}-\ref{15})
have been determined using the kernel $\tilde M(x,y)$, given by Eqs.
(\ref{22}) and (\ref{23}).  Some results are shown in Table 1 and
Fig.~2.  In our present study we have taken $m=0$.  Note, that
$\tilde M(x,y)$ is approaching a local linear potential at large
distances, $x,y\ga \sigma^{-1/2}$, which justifies a posteriori our
choice of the reference basis. Due to the nonlocality of the
interaction the predicted spectrum is found to be different from that
of the linear potential, valid at large distances.  Moreover,
comparing the level structures of the $J=\frac{1}{2}$ channel as
obtained using the kernels (\ref{22}) and (\ref{23}) we see from
Table I, that they are qualitatively very similar, corroborating that
there is indeed a strong forward focussing effect in the quark
propagator.  From Fig. 2 we see that for all L values the higher
radially excited levels are close to the
predictions of  linear potential $V_0(r)=
0.9 \sigma r - 1.8 \sigma^{1/2}$, in agreement with the
fact that the interaction at large distance can indeed be described
by a local linear potential. On the other hand the nonlocal kernel
predictions for the low lying states clearly deviates strongly from
those of the (shifted) linear potential. Hence the nonlocal nature of
the force does affect the spectrum in an essential way.

The eigenfunctions for the nonlocal kernels look qualitatitively
similar to the corresponding ones of the shifted linear potential.
In Fig.~3 are shown the ground state and first excited state
for the $J=\frac{1}{2}$ channels. Altough the differences are
substantial for these low lying states, the agreement for higher excited
states is considerably better. Moreover, we find that the large
distances and high  states of the WKB states agree well with the
corresponding eigenfunctions.

\subsection{Inclusion of perturbative exchanges}

Till now we have considered only NP part of the gluonic field, $B_\mu$.
In this section we include the perturbative part, $a_\mu$, and
neglect for simplicity the interference terms. Therefore the effect  of
$a_\mu$ is accounted for in the appearance of an additional factor in
the partition function (\ref{1}), namely
\be
Z=Z_{NP} Z_{pert},\ \ Z_{pert}=\langle e^{g \int dx \psi^+\hat a(x) \psi
(x) + i g \int dz_4 a_4 (z_4)}\rangle_a,
\label{24}
\ee
where the second term in the exponent of (\ref{24}) corresponds to the 
interaction of the perturbative part of the gluon field with the
static antiquark. We have used in (\ref{24}), that due to the 'tHooft
identity \cite{18} one can average independently over $B_\mu$ and
$a_\mu$.

The result of averaging yields a new additive term in $L_{eff}$, 
Eq.~(\ref{1}),
  \be
  L_c=g\int dx \psi^+ (x) \hat A^{(c)}(x) \psi(x),
\label{25}
  \ee
where we have defined
  \be
  A_{\mu}^{(c)}(x)=-ig \int dz_4 <a_4(x) a_4(z_4)> =\delta_{\mu 4}
  \frac{(-i)gC_2}{4\pi|\mbox{\boldmath ${\rm x}$} |}.
\label{26}
  \ee
  The presence of $L_c$ in Eq.~(\ref{1}) does not influence the derivation
  of basic Eqs.~(\ref{9}-\ref{10}). The only difference is
  that Eq.  (\ref{10}) assumes the form
   \be (-i\hat \partial -g
  \hat A^c(x)-im) S(x,y)-i\int M(x,z) S(z,y) d^4z=\delta^{(4)}(x-y).
  \label{27}
  \ee
Eq. (\ref{9}) does not change  and the kernel
  $J(x,y)$ contains as before only nonperturbative
  contributions. Note however that $S(x,y) $ in $M(x,y)$ in
Eq. (\ref{9}) now contains also perturbative gluon exchanges.
  This is a new type of interference of perturbative and NP terms,
  which appears irrespectively of our neglect of this interference
  within the averaging procedure over $B_\mu$ and $a_\mu$. In other
  words another class of diagrams is responsible for this
  interference.

  Correspondingly in the static
   equations (\ref{11}) one should replace
   \be
   \beta m \to \beta m -\frac{C_2\alpha_s}{|\mbox{\boldmath ${\rm
   z}$}|}
   \label{28}
   \ee
The equations for the partial waves (\ref{14}-\ref{15}) are modified
   due to the presence of the color Coulomb potential $V(r)$ in a
   simple way. Since
   \be
    V(r)=-\frac{C_2\alpha_s}{r}
    \label{29}
    \ee
     is local and a
   Lorentz vector, it always appears in the combination
   $\varepsilon_n-V(r)$. Hence one has instead of
Eqs.~(\ref{14}-\ref{15})
   \be
\frac{dF_\nu}{dr}-\frac{\kappa}{r}F_\nu+(\varepsilon_\nu-V(r)-m)
G_\nu- M_{11} G-i M_{12} F=0,
\label{30}
\ee
\be
\frac{dG_\nu}{dr}
+\frac{\kappa}{r}G_\nu-(\varepsilon_\nu-V(r)+m)
F_\nu- M_{22} F+i M_{21} G=0,
\label{31}
\ee
where  we have denoted
\be
M_{ik}
\left(\begin{array}{l}
G\\
F
\end{array}
\right) \equiv\int <\nu|M_{ik}|\nu'>
\left(\begin{array}{l}
G_{\nu'}(w)\\
F_{\nu'}(w)
\end{array}
\right)
rwdw.
\label{32}
\ee
Here $M_{ik}$ is defined as in Eq. (\ref{12}) and the matrix
$\Lambda_{ik}$ in Eq. (\ref{12}) involves the sum over all states,
including positive and negative $\varepsilon_n$. There in section IIB
we have exploited the symmetry (\ref{17}). However 
Eqs.~(\ref{30}-\ref{31}) are invariant under another
transformation, namely \be \varepsilon_n\leftrightarrow
-\varepsilon_n, V(r)\leftrightarrow -V(r), \kappa\leftrightarrow
\kappa, G_n\leftrightarrow F_n.  \ee Now the sum over negative
  $\varepsilon_n$ can be expressed through the corresponding sum over
positive $\varepsilon_n$  with exchange $G_n\leftrightarrow F_n$ as
before, but also with the inversion of sign of Coulomb interaction,
 i.e. Coulomb attraction for positive $\varepsilon_n$ is replaced by
 Coulomb repulsion for negative $\varepsilon_n$.

 In what follows we shall  denote  wave functions of the positive
 energy states with repulsive Coulomb with  the sign of tilde:
 $\tilde G_\nu, \tilde F_\nu$.
 Then using (\ref{12}) the matrix $\beta \Lambda_{ik} $ can be
 written as a sum over  only positive $\varepsilon_n$ as follows
 \be
 \beta\Lambda^{\mu\mu'}_{ik} = \frac{1}{xy}\sum_{jlM,n>0}
\left(\begin{array}{ll}
~~~~G_\mu G_{\mu'}^*- \tilde F_{\mu} \tilde F^*_{\mu'},
&-i(G_{\mu}F^*_{\mu'}- \tilde F_\mu\tilde G^*_{\mu'})\\
-i(F_\mu G_{\mu'}^*- \tilde G_{\mu} \tilde F^*_{\mu'}),
&~~~~\tilde G_{\mu}\tilde G^*_{\mu'}-  F_\mu F^*_{\mu'}
 \end{array}
\right).
\label{34}
\ee
Since $\beta\Lambda$ is  exactly the combination which enters the mass
matrix (\ref{12}), one can list in (\ref{34}) scalar and vector
(proportional to $\beta)$ parts:
 \be M=M_s I
+M_v \beta +\Delta M,
 \label{35}
 \ee
  where $\Delta M$ contains
spin-dependent terms, which can be considered as in section IIB,
 while $M_s$, $M_v$ are
 \be
M_{s,v}=C\sum_{jlM\mu, n>0}
[G_\mu G^*_{\mu}-\tilde F_\mu\tilde F^*_\mu\pm (\tilde G_\mu \tilde
G_\mu - F_\mu F^*_\mu)]
\label{36}
\ee
where
\be
 C=\frac{1}{4} \sqrt{\pi} T_g D(0) \frac{\mbox{\boldmath
${\rm  x}$}\cdot \mbox{\boldmath ${\rm y}$}}{xy}
f(\mbox{\boldmath ${\rm x}$}, \mbox{\boldmath ${\rm y}$}),
~~f({\bf x},{\bf y}) =\int^1_0 ds \int^1_0 dt ~exp\left(-\frac{( {\bf
x}s-{\bf y }t)^2}{4T^2_g}\right).
 \label{37}
 \ee
  From (\ref{36}) it is clear that the vector part $M_v$ is only due
to the presence of Coulomb interaction.
Corrections at large distances due to the vector part can be treated again in
the relativistic  WKB.  A rough estimate of $M_v$ at large $r$ yields
\be
\frac{M_v}{M_s}\sim \frac{ \alpha_s}{\sigma r^2}
\label{38}
\ee
and hence can be neglected at large enough $r$.

\section{Numerical solutions of equations and comparison to $B$, $D$
mesons}

We have performed numerical studies of Eqs. (\ref{30}-\ref{31}) with
the kernel (\ref{22}) for different values of the quark mass $m$ and
different values of $T_g$. To simplify calculations only the dominant
part of the mass operator $ M_{ik}$ was retained, i.e. $\hat
M_{11}= M_{22}= M^{(0)}$, while $ M_{12},  M_{21}$ have
been neglected.  For  $M^{(0)}$ the representation (\ref{kern}) was
used
$$
M^{(0)}({\bf x},{\bf y}) = J({\bf x,} {\bf y}) \tilde \Lambda(
{\bf x},{ \bf y}) I,
$$
where $\tilde \Lambda$ is taken to be the kernel (\ref{22}).
  Results of our calculations for the ground state energy are listed
in Table 2. One can see a rather sharp change of energy when $\alpha_s$
changes from 0 to 0.3 and when $T_g$ is changing from 0 to 0.25,
while further increase of $\alpha_s$ or $T_g$ does not produce such
a strong dependence.

 Solutions of our equations (\ref{30}-\ref{31}) can
be compared with physical states of $B$, $D$ and $B_s, D_s$ mesons.
To this end one should have in mind that in Eqs. (\ref{30}-\ref{31}) the
static approximation for the heavy quark $b,c$ was used,  and hence
all corrections $O(1/m^n_Q)$ with $n\geq 1$ are neglected.

One can exploit at this point the HQET expansion for the mass $m_H$
of heavy--light boson \cite{19,20}
\be
m_H=m_Q(1+\frac{\bar \Lambda}{m_Q}+\frac{1}{2m^2_Q}
(\lambda_1+d_H\lambda_2) +O(1/m^3_Q ),
\label{39}
\ee
where $\lambda_n$ are free parameters, depending on dynamics,
and $d_H$ is the hyperfine splitting parameter.
It is clear  from the preceding that eigenvalues of Eqs.
(\ref{30}-\ref{31}) yield the value $\bar \Lambda$, which depends
on the quantum numbers of the state,
\be
\bar \Lambda (j,l,n_r) =\varepsilon _n(j,l)
\label{40}
\ee

Consider now the results of the present approach, i.e. solutions of
Dirac--type equations (\ref{30}-\ref{31}).
In the local case $(T_g\to 0)$ when the kernel $ M$ reduces to
the linear potential $\sigma r$, we have
\be
\bar \Lambda_D^{(loc)} = 0.690 ~GeV ~~(\alpha_s=0, ~~\sigma =0.18 ~GeV^2)
\label{41}
\ee
and
\be
\bar \Lambda_D^{(loc)} = 0.493 ~GeV ~~(\alpha_s=0.3, ~~\sigma =0.18
~GeV^2).
 \label{42}
  \ee
 This should be compared to the nonlocal case
\be
\bar \Lambda_D^{(nonloc)} = 0.415 ~GeV ~~(\alpha_s=0, ~~\sigma =0.18
~GeV^2)
 \label{43}
  \ee
and
   \be
   \bar \Lambda_D^{(nonloc)} = 0.288~GeV ~~(\alpha_s=0.3, ~~\sigma =0.18 ~GeV^2).
  \label{44}
   \ee
  These latter values are in  general agreement with the results of
  the QCD heavy--flavour sum rules \cite{BSUV,BB}
  \be
   \bar \Lambda= 0.57\pm 0.07 ~GeV
   \label{45}
   \ee
   and more recent analysis from  semileptonic $B$ decays \cite{21}
  \be
   \bar \Lambda= 0.39\pm 0.11 ~GeV
   \label{46}
   \ee

   Another interesting comparison is with the experimental values of
  the  $B$--meson mass (the term $\lambda_2$ in Eq.~(\ref{39}) can 
  be determined from the $B^*-B$ mass difference).
   Using Eq. (\ref{44}) and  $\bar M_B=\frac{3M_{B^*}+ M_{B}}{4} =
5.312 ~GeV$ one can estimate (neglecting $\lambda_1)$ the pole mass of
   the $b$--quark  to be $m_b(pole)\cong 5.0 ~GeV$, which is  in
   reasonable agreement with the analysis of the quarkonium spectra in
   \cite{22}.

   A similar analysis can be done for the $B_s$ meson; the
   corresponding values for $\bar \Lambda_s$ with $m_s=0.15 $ and
   0.20 $GeV$, are
    $\bar \Lambda_s-\bar \Lambda =0.084$ and
    $\bar \Lambda_s-\bar \Lambda =0.115$ for $\alpha_s=0.3$ .
    One can compare these values with
     the mass difference $B_s, B^0,
     \Delta M_s(B)=(0.090\pm 0.0038)
     ~GeV$.
These numbers for $\bar \Lambda$ can be compared with those in Table
3, where also results of lattice calculations \cite{GMS} and of
the constituent quark model (CQM) \cite{F,HKN,27} are
given.

\section{Chiral condensate}

As a  check of CSB in our Eqs.~(\ref{14}-\ref{15}) we have
computed the chiral condensate, which can be expressed through the
eigenfunctions as in \cite{S2} (to simplify matter we disregard in
this section perturbative contributions).
  \be
   \langle \bar q
 q\rangle =-\frac{N_c}{2\pi}\sum^{\infty}_{n=0}[(A^-_n)^2-(B^+_n)^2],
\label{47}
\ee
where
$A^-_n=(\frac{G_n(r)}{r})_{r=0},B^+_n=(\frac{F_n(r)}{r})_{r=0}$,
and $G_n,F_n$ refer to solutions with $\kappa=-1, l=0$ and
$\kappa=+1, l=1$ respectively.
In the local linear potential case the values of $A^-_n, B^+_n$ have
been computed in the WKB method \cite{S2} and shown to yield a
monotonically divergent series $\langle \bar q
q\rangle =-\frac{N_c}{2\pi}\sum_n\frac{const}{\sqrt{n}}$.

It can be argued (using Eqs. (8-9) from \cite{S2}), that the
nonlocality of the kernel $ M$ in space--time, present by
definition in Eq. (8) improves the convergence of the series and yields a
finite result for $\langle \bar q q\rangle $.
We have found $A_n^-, B_n^+$ from the solutions of the nonlocal
equations (\ref{14}-\ref{15}) with the kernel $\tilde M$  and
compared them with the local case, when $\tilde M$ reduces to the
local linear potential. Results are shown in Table 4.

One can see from the results, that in the nonlocal
case the magnitude of
$s_n\equiv (A^{-}_n)^2-(B^{+}_n)^2$, is
clearly  diminished as compared to the
reference local case, and is of reasonable order of magnitude. From 
the obtained sequence of $s_n$ we get that
$\langle \bar q q\rangle=-0.5 \sigma^{3/2}$ and $-0.7 \sigma^{3/2}$
in the nonlocal cases of the kernels (21) and (22) respectively.
Adopting a value of $\sigma=0.2 ~GeV^2$ we find
$\langle \bar q q\rangle=-(350 ~MeV)^3$
and $-(400 ~MeV)^3$ respectively, to be compared with the
usually acceptable value of $-(250 ~MeV)^3$.
However convergence is still slow as seen from Table 4 and the
converged values are somewhat higher. We have checked that
convergence is somewhat improved when one takes into
account the intrinsic nonlocality of the kernel $ M$ in $\vex,
\vey$. To this end we have modified the kernel $\tilde M$ obtained from WKB analysis, 
replacing $\tilde \delta$ in Eq. (21) by a Gaussian factor
\be
N exp (-\frac{(x-y)^2}{a^2}) \delta(cos \theta_x -cos\theta_y)
\label{48}
\ee
and studied the sequences of
$s_n$ as functions of the nonlocality range $a$. Results are shown in
Table 4 for two values of $a=0.3 \sigma^{1/2}$ and $0.5 \sigma^{1/2}$.
The strength $N$ is chosen such that
numerically the slope of $\sigma=0.2~GeV^2$ is reproduced for large distances.
The condensate values varies in the considered region substantially, showing
that effects of lonlocality are important.

The slope of the effective potential $V_{eff}(r)$, determined by $N$,
strongly depends on $a$   for $a\cong T_g$.
We believe that the reason for this lies in the fact,
that the chiral condensate $\langle \bar q q\rangle$ depends crucially on
the nonlocality both in time components of $ M(\vex, \vey;x_4, y_4)$
and in spacial components. The first nonlocality was however
disregarded in Eqs.~(\ref{11}-\ref{12}), when the $p_4$ --
dependence was omitted in $ M$ and $\psi_n$ (the static limit).
It was indeed shown in Ref.~\cite{S2}, that taking this dependence into
account significantly improves convergence of the sum in Eq.
(\ref{47}).  The full account of this effect requires solution of
time--dependent Eqs.~(\ref{9}-\ref{10}), which is numerically  a
much more difficult problem.

\section{Conclusion}

We have studied the confining and CSB properties in the
system of one light quark and one static antiquark.
The effective mass operator is written explicitly for large $N_c$, as
 a sum  over vacuum field correlators. Keeping only the Gaussian field
correlator, we
have obtained a closed system of equations in the limit of large
$N_c$.
Our results support the presence of a Lorentz scalar linear confinement for
the light quark, which signifies CSB for this system, and yield
eigenfunctions and eigenvalues for the heavy--light system containing
both confinement and CSB.

As a direct evidence of CSB we have computed the chiral condensate,
which appears to be of the correct sign and having the proper large
$N_c$ dependence.
Our result yields a reasonable order of  magnitude of $\langle \bar
q q\rangle $, provided convergence of the sum is achieved.
At this point it is useful to compare the CSB picture of the NJL model and our
approach. In the NJL model confinement and string are absent and CSB
may occur due to the condensation of $q {\bar q}$ pairs in the
scalar channel. In our case, being the large $N_c$ approximation of the real
QCD, a string is built up between light and heavy quark, which
depends not only on light quark coordinates $\vex, \vey$, but also
on the distance from them to the heavy antiquark.
In the presence of confinement,
the phenomenon of CSB is due to the spontaneous creation of
the scalar string, which is forbidden by chiral symmetry.

Eigenvalues $\varepsilon_n$ and eigenfunctions obtained
numerically for lowest states, represent the leading
contributions of the HQET expansion in powers of $1/m_Q$.
Results for the energies $\varepsilon_n$ in our method are
compared of the lattice and QCD sum rule calculations, and also
with experimental extraction of $\varepsilon_n=\bar \Lambda
(n)$, showing an overall agreement with the B and D meson masses.

\vspace{0.5cm}
\noindent
{\bf Acknowledgement}

This work was started while one of the authors (Yu.S.) was a
visitor at ITP Utrecht. The hospitality of the Institute and
financial support by FOM are gratefully  acknowledged.
Yu.S. is grateful to I.M.Narodetsky for a useful discussion.

\newpage

\epsfxsize=12cm
\begin{figure}[h]
\begin{center}
\epsffile{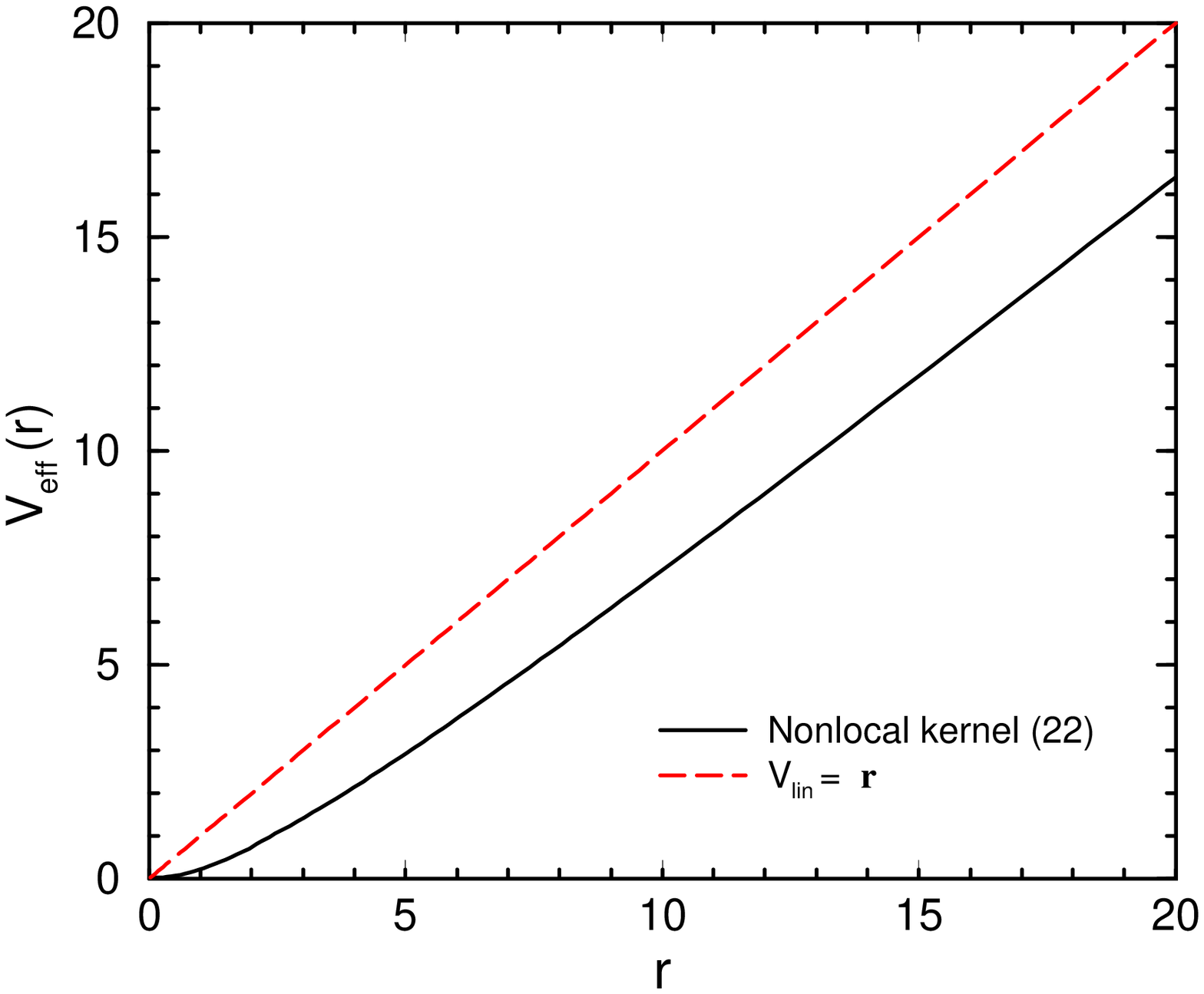}
\end{center}
\caption{Behaviour of the effective potential,
$V_{eff}(r)=\int\tilde M_{11}(r,x)rxdx$, as a function of
$r$ (solid line). The kernel (\ref{22}) and units of $\sqrt{\sigma}$ have been used.
For comparison, the linear local potential $V_{lin}(r)=
r$ (dashed line) is also plotted.}
\figlab{effpot}
\end{figure}

\epsfxsize=12cm
\begin{figure}[h]
\begin{center}
\epsffile{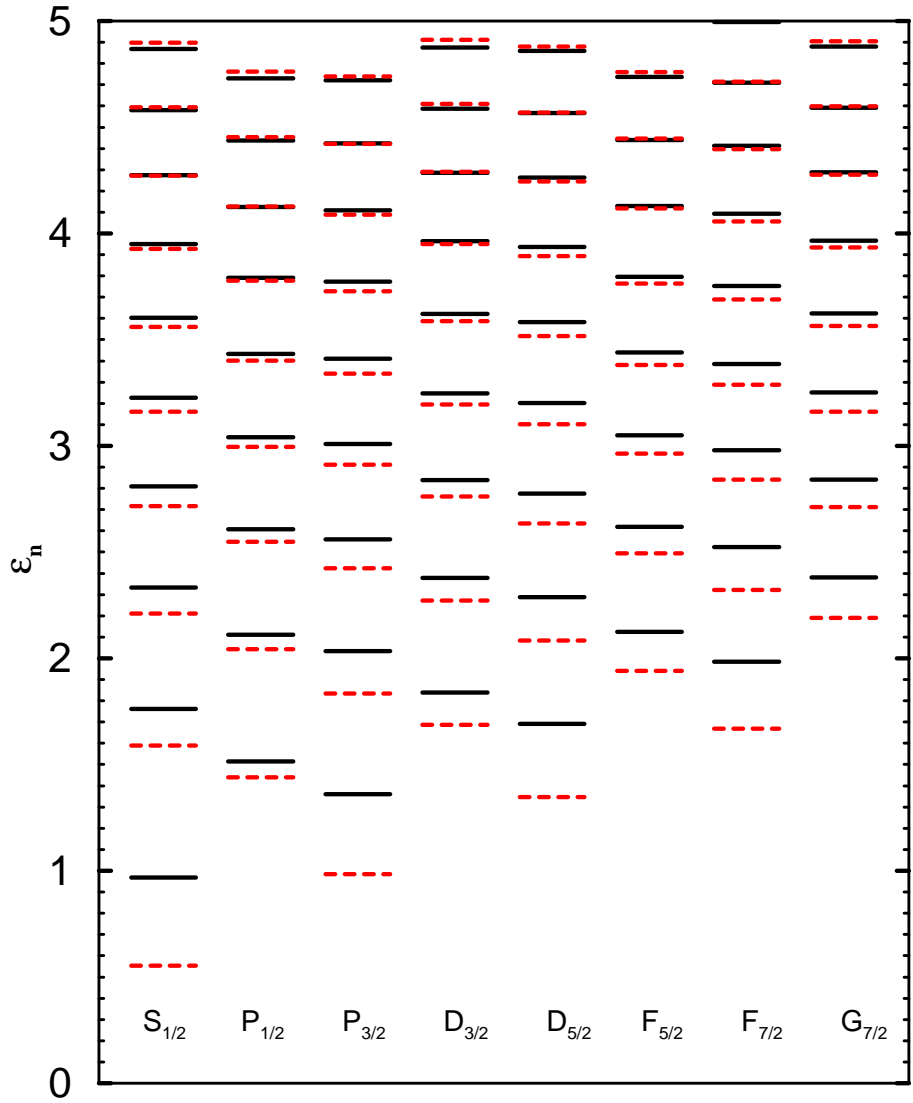}
\end{center}
\caption{
Level Structure calculated with the Dirac equation using  the
kernel (\ref{23}) (solid line)
in comparison with the predictions of the linear potential
$V_0(r)=0.9 r - 1.8$ (dashed line).
States carry the quantum numbers $L_J$, with $L,J$ being the orbital and total
angular momentum.}
\figlab{spectrum}
\end{figure}

\epsfxsize=12cm
\begin{figure}[h]
\begin{center}
\epsffile{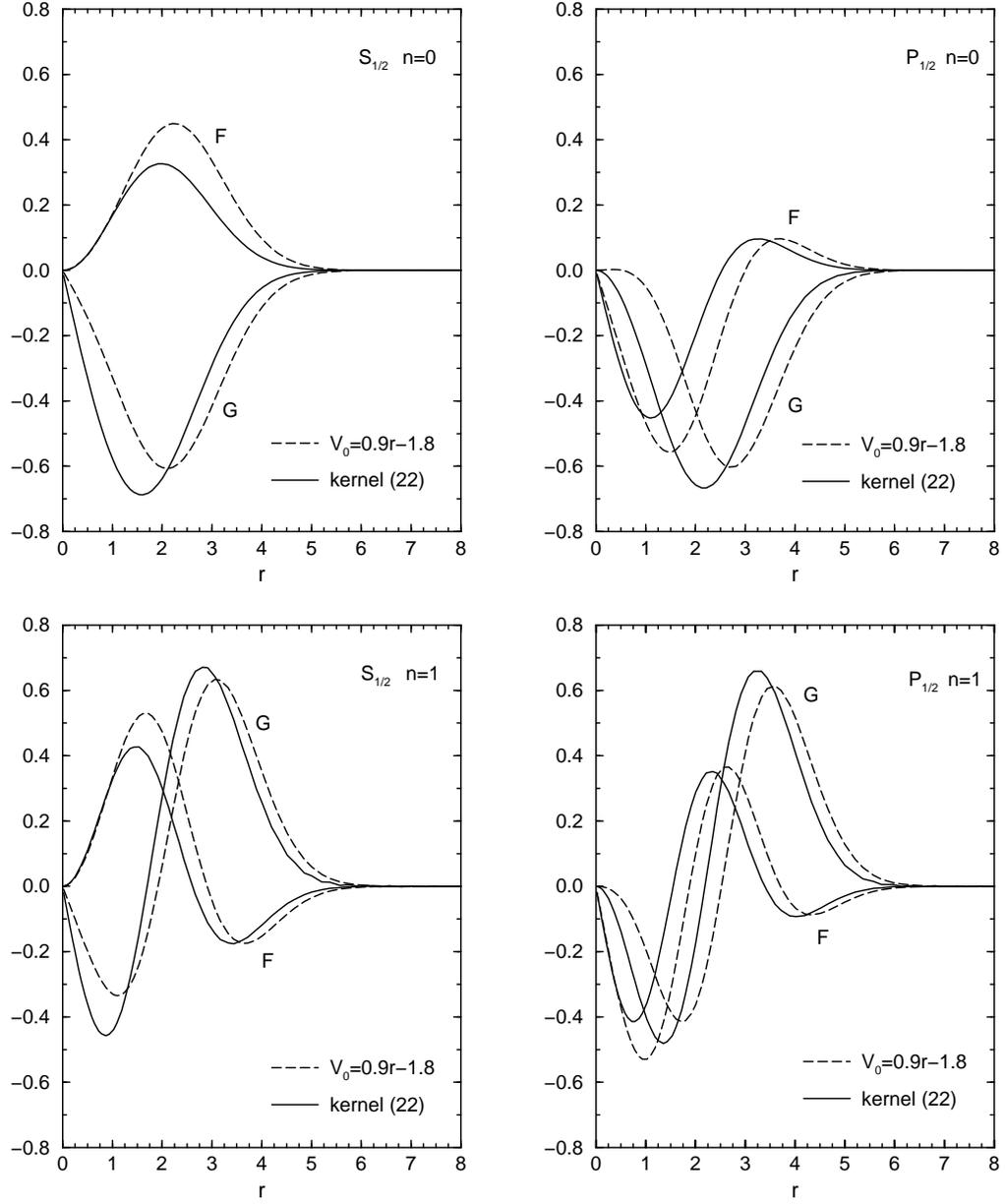}
\end{center}
\caption{ Eigenfunctions of the ground state and the first excited
state for the $S_{\frac{1}{2}}$ and $P_{\frac{1}{2}}$ channels.
The solutions correspond to using  the kernel (\ref{23}) (solid line)
and the shifted linear potential
$V_0(r)=0.9 r - 1.8$ (dashed line).}
\figlab{wavef}
\end{figure}

\newpage
Table 1.   Energy eigenvalues for Eqs. (13-14) with $J=\frac{1}{2}$
and the kernels, given by Eq. (\ref{22}) and Eq. (\ref{23}), as compared
with those of the Dirac equation for the linear local potential
$ V_{lin}(r)= r$.
\begin{center}
\begin{tabular}{|l|l|l|l|l|l|l|}  \hline
	& \multicolumn{2}{c|}{$V= r$}
	& \multicolumn{2}{c|}{kernel (\ref{22})}
	& \multicolumn{2}{c|}{kernel (\ref{23})}\\
\cline{2-7}
n	&$\kappa=-1$  & $\kappa=1$ & $\kappa=-1$ & $\kappa=1$ &$\kappa=-1$  & $\kappa=1$ \\
\hline
0	& 1.619	& 2.294 & 0.925  & 1.472 & 0.969 & 1.516  \\
1	& 2.603	& 3.031 &  1.719 & 2.056 & 1.765 & 2.113 \\
2	& 3.291	& 3.626 & 2.277  & 2.541 & 2.334 & 2.608 \\
3	& 3.855	& 4.138 & 2.740  & 2.964 & 2.809 & 3.042 \\
4	& 4.345	& 4.594 & 3.144  & 3.342 & 3.226 & 3.432 \\
5	& 4.784	& 5.008 & 3.507  & 3.685 & 3.603 & 3.790 \\
6	& 5.186 & 5.334 & 3.838  & 4.002 & 3.950 & 4.123 \\
\hline
\end{tabular}
\end{center}
~    \\

Table 2.  Ground state energy eigenvalue (in units of
$\sqrt{\sigma}$) for Eqs. (\ref{30}-\ref{31}) with
$\alpha_s=0$, 0.3 and 0.39 and quark masses $m=5 ~MeV$, $0.15 ~GeV$ and
$0.2 ~GeV$ (upper, middle and lower entry) for different values of
$T_g$, $T_g=0$, 0.25, 0.5 and 1 (in units of $1/{\sqrt{\sigma}}$).

\begin{center}
\begin{tabular}{|ll|l|l|l|l|}  \hline
& $T_g$& 0& 0.25& 0.5& 1\\
$\alpha_s$ &&&&&    \\\hline
&&1.628& 0.985& 0.979& 0.907\\
0&& 1.886& 1.225& 1.217& 1.145\\
&& 1.978&1.314&1.305&1.233\\\hline
&&1.163&0.684&0.679&0.628\\
0.3&&1.378&0.884&0.877&0.826\\
&&1.456&0.959&0.951&0.900\\\hline
&&1.004&0.585&0.580&0.536\\
0.39&&1.201& 0.768&0.761&0.717\\
&&1.272&0.837&0.830&0.786\\
\hline
\end{tabular}
\end{center}

      \newpage
Table 3. Energy eigenvalues $\bar \Lambda$ of the heavy--light system
in the static heavy quark approximation  obtained in  different
approaches.\\

\begin{center}
\begin{tabular}{|l|l|l|}
\hline
Refs.& Method &$\bar \Lambda ~(GeV)$\\ \hline
20 & QCD sum rules& $>0.5$\\
21 & QCD sum rules& $0.4\div 0.5$\\
24 &Lattice& $0.18\pm 0.03$\\
22 &experim. & $0.33\pm 0.11$\\
25 &QCM  & $0.35$\\
26 &QCM  & $0.5\div 0.6$\\
27 &Rel. QCM  & $0.386$\\
this work &Nonlin. Dirac Eq.   & $0.287$\\ \hline
\end{tabular}
\end{center}

\vspace{1cm}
Table 4. The difference $s_n= |A_n|^2-|B_n|^2$ for $n=0,1,... 6$, in
case of the nonlocal kernels (\ref{22}) and (\ref{23}) and corrected for a normalized
Gaussian nonlocality ({\ref{48}) with a range of $a$ =0.3 and $a$=0.5.
For comparison the results are shown for a local linear potential
$V_{lin}= r$ .\\

\begin{center}
\begin{tabular}{|l|l|l|l|l|l|}
\hline
	& \multicolumn{5}{c|}{$A_n^2-B_n^2$}\\
\cline{2-6}
n	& $V= r$& kernel (\ref{22}) & kernel (\ref{23}) & $a$=0.3 & $a$=0.5 \\
\hline
0 & 0.79 & 0.42 & 0.50 & 0.15 & 0.23
\\
1 & 0.51 & 0.21 & 0.34 & 0.04 & 0.12
\\
2 & 0.41 & 0.12 & 0.19 & 0.02 & 0.10
\\
3 & 0.35 & 0.10 & 0.16 & 0.02 & 0.09
\\
4 & 0.31 & 0.08 & 0.11 & 0.01 & 0.09
\\
5 & 0.27 & 0.07 & 0.11 & 0.01 & 0.08
\\
6 & 0.26 & 0.06 & 0.09 & 0 & 0.07
\\
\hline
\end{tabular}
\end{center}

  \end{document}